# Software Engineering with Process Algebra: Modelling Client / Server Architectures

*Bob Diertens*

Programming Research Group, Faculty of Science, University of Amsterdam

*ABSTRACT*

In previous work we described how the process algebra based language PSF can be used in software engineering, using the ToolBus, a coordination architecture also based on process algebra, as implementation model. We also described this software development process more formally by presenting the tools we use in this process in a CASE setting, leading to the PSF-ToolBus software engineering environment. In this article we summarize that work and describe a similar software development process for implementation of software systems using a client/server model and present this in a CASE setting as well.

## 1. Introduction

In previous work [9-11] we investigated the use of process algebra, in particular the process algebra based language PSF (Process Specification Formalism), in the software development process. We used the ToolBus, a process algebra based coordination architecture, as implementation model. We give a description of PSF and the ToolBus in sections 1.1 and 1.2. We described this software development process more formally in [12] by presenting the tools we use in the development process in a Computer-Aided Software Engineering Environment (CASE) setting.

In the work mentioned above we only used the ToolBus as target system model for implementing software systems. Our goal is to support software development with process algebra for other system models as well. In this article we describe how client / server based architectures can be developed using process algebra. Furthermore, we describe this process in a CASE setting leading to a software engineering environment.

In client / server architecture based software systems tasks are partioned between service providers (servers) and service requesters (clients). Such software systems consist of one or more clients that make use of services provided by one or more servers. But a server itself can also act as a client making requests to one or more other servers. In this way, a hierarchy is formed in which clients and servers on a lower level can make requests to servers on higher levels.

In the remainder of this section we give brief descriptions of PSF and the ToolBus. In section 2 we describe our software engineering process with PSF and the software engineering environment supporting this process. In section 3 we show how the modelling of client / server architectures in PSF can be achieved and in section 5 we describe this process more formally by presenting it in a CASE setting, leading to the PSF-Client / Server Software Engineering Environment. We briefly discuss the implementation of applications based on the specifications of client / server architectures in section 5. We end with sections on related work and conclusions.

*1.1 PSF*

PSF is based on ACP (Algebra of Communicating Processes) [3] and ASF (Algebraic Specification Formalism) [4]. A description of PSF can be found in [13, 14, 20, 21]. Processes in PSF are built up from the standard process algebraic constructs: atomic actions, alternative composition +, sequential composition ·, and parallel composition ∥ . Atomic actions and processes are parameterized with data parameters.

PSF is accompanied by a Toolkit containing among other components a compiler and a simulator that can be coupled to an animation [15]. The tools operate around the tool intermediate language (TIL) [22]. Animations can either be made by hand or be automatically generated from a PSF specification [16]. The animations play an important role in our software development process as they can be used to test the specifications and are very convenient in communication to other stakeholders.

*1.2 ToolBus*

The ToolBus [5] is a coordination architecture for software applications developed at CWI (Amsterdam) and the University of Amsterdam. It utilizes a scripting language based on process algebra to describe the communication between software tools. A ToolBus script describes a number of processes that can communicate with each other and with various tools existing outside the ToolBus. The role of the ToolBus when executing the script is to coordinate the various tools in order to perform some complex task. A language-dependent adapter that translates between the internal ToolBus data format and the data format used by the individual tools makes it possible to write every tool in the language best suited for the task(s) it has to perform.

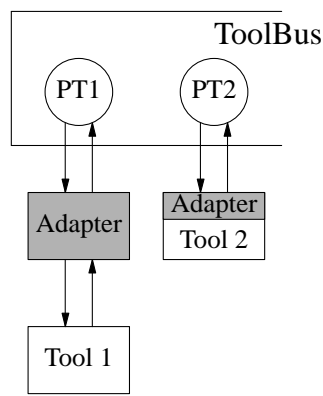

**Figure 1.** Model of tool and ToolBus interconnection

In Figure 1 two possible ways of connecting tools to the ToolBus are displayed. One way is to use a separate adapter and the other to have a built-in adapter. Processes inside the ToolBus can communicate with each other using the actions `snd-msg` and `rec-msg`. ToolBus processes can communicate with the tools using the actions `snd-do` and `snd-eval`. With the latter a tool is expected to send a value back which can be received by the process with the action `rec-value`. A tool can send an event to a ToolBus process which can be received with the action `rec-event`. Such an event has to be acknowledged by the ToolBus process with the action `snd-ack-event`.

**2. Software Engineering with PSF**

In [12], previous work on software engineering with PSF is summarized and put in a CASE setting, resulting in a software engineering environment based on process algebra. In this section we briefly describe this environment and give an example of its use.

*2.1 The PSF-ToolBus Software Engineering Environment*

In Figure 2 we show the PSF-ToolBus software engineeringing environment that can be used for the development of ToolBus applications. Objects to be specified are presented as **bold boxes**, workbench tools as ellipses, and generated objects as *slanted boxes*. The environment consist of two workbenches, one for the specification of the architecture of the software system, and one for the specification of the software system as ToolBus application. Each workbench uses a library of PSF modules in which the primitives for this particular abstraction level are specified. On each level the connection of the components into a system and the incorporation in an environment is generated from the components.

The development of a software system starts with the specification of the architecture of the software systems from which a ToolBus application specification can be obtained by applying vertical and horizontal implementation techniques based on our process algebra. Vertical implementation is the refining of actions in the architecture specification by mapping these actions onto sequences of actions. Horizontal implementation is the constraining of the processes that are the result of the vertical implementation with processes that specificy the tools. A process can be constrained by another process by putting the processes in parallel with each other and enforcing communication between the two by encapsulation.

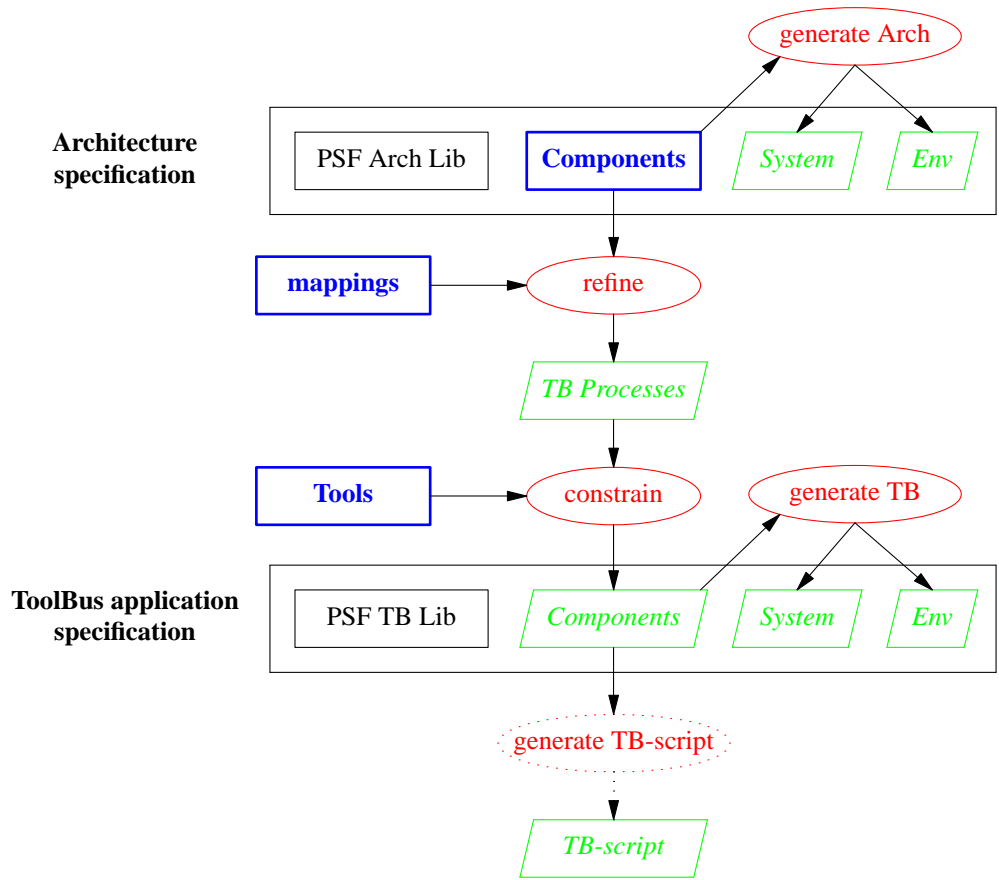

**Figure 2.** The PSF-ToolBus Software Engineering Environment

From the specification of the ToolBus processes in the ToolBus application specification a ToolBus script can be derived that together with the implementation for the tools form a ToolBus application. The derivation of the ToolBus script is not done automatically. The problem here is that PSF specifications use recursion for setting the state of a process, and the ToolBus cannot handle recursive processes.

*2.2 Example*

We show our development process for a small application. In this example, Component1 can either send a `message` to Tool2 and then wait for an acknowledgement from Component2, or it can send a `quit` after which the application will shutdown.

**Architecture Specification**

We first specify a module for the data and id's we use.

```
data module Data
begin
    exports
    begin
```

```
        functions
            message : → DATA
            ack : → DATA
            quit : → DATA
            c1 : → ID
            c2 : → ID
        end
    imports
        ArchitectureTypes
end Data
```

We then specify the system of our application.

```
process module ApplicationSystem
begin
    exports
    begin
        processes
            ApplicationSystem
    end
    imports
        Data,
        ArchitecturePrimitives
    atoms
        send-message
        stop
    processes
        Component1
        Component2
    definitions
        Component1 =
            send-message .
            snd(c1 >> c2, message) .
            rec(c2 >> c1, ack) .
            Component1
          + stop .
            snd-quit
        Component2 =
            rec(c1 >> c2, message) .
            snd(c2 >> c1, ack) .
            Component2
        ApplicationSystem = Component1 ∥ Component2
end ApplicationSystem
```

The `snd-quit` in the process definition for Component1 communicates with the architecture environment followed by a disrupt to end all processes.

Next, we put the system in the architecture environment by means of binding the main process to the System parameter of the environment.

```
process module Application
begin
    imports
        Architecture {
            System bound by [
                System → ApplicationSystem
            ] to ApplicationSystem
            renamed by [
                Architecture → Application
            ]
        }
end Application
```

The generated animation of the architecture is shown in Figure 3. Here, Component1 has just sent a message to Component2, which is ready to send an acknowledgement back. Each box represents an encapsulation of the processes inside the box, and a darker ellipse is a process which is enabled to perform an action in the given state.

The module mechanism of PSF can be used to build more complex components hiding internal actions and sub-processes. With the use of parameterization it is even possible to make several instances of a component.

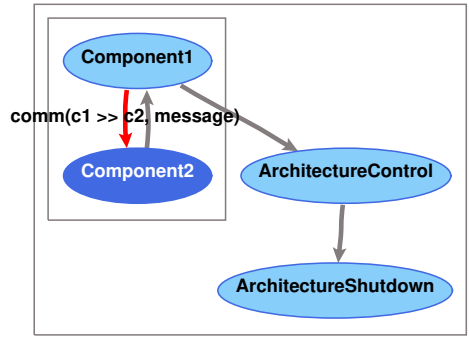

**Figure 3.**  Animation of an example architecture

**ToolBus Application Specification**

We make a ToolBus application specification for our example in the form shown in Figure 1.  By refining the specification of the architecture we obtain a ToolBus application specification for our example.  Take the process `Component1` from the architecture specification of our toy example.

```
Component1 =
    send-message .
    snd(c1 >> c2, message) .
    rec(c2 >> c1, ack) .
    Component1
  + stop .
    snd-quit
```

We can make a virtual implementation by applying the mapping consisting of the refinements

```
snd(c1 >> c2, message) → tb-snd-msg(t1, t2, tbterm(message))
rec(c2 >> c1, ack)     → tb-rec-msg(t2, t1, tbterm(ack)) .
                         tb-snd-ack-event(T1, tbterm(message))
snd-quit               → snd-tb-shutdown
```

and the renamings of the local actions

```
send-message         → tb-rec-event(T1, tbterm(message))
stop                 → tb-rec-event(T1, tbterm(quit))
```

Renaming the process `Component1` into `PT1` gives the following result.

```
PT1 =
    tb-rec-event(T1, tbterm(message)) .
    tb-snd-msg(t1, t2, tbterm(message)) .
    tb-rec-msg(t2, t1, tbterm(ack)) .
    tb-snd-ack-event(T1, tbterm(message)) .
    PT1
  + tb-rec-event(T1, tbterm(quit)) .
    snd-tb-shutdown
```

We can show that `Component1` and `PT1` are vertical bisimular.  Applying the renamings on process `Component1` and hiding of the actions to be refined results in

```
Component1' =
    tb-rec-event(T1, tbterm(message)) . τ . τ . Component1'
  + tb-rec-event(T1, tbterm(quit)) . τ
```

Hiding of the actions in the refinements in process `PT1` results in

```
PT1' =
    tb-rec-event(T1, tbterm(message)) . τ . τ . τ . PT1'
  + tb-rec-event(T1, tbterm(quit)) . τ
```

It follows that `Component1'` and `PT1'` are rooted weak bisimilar.

We now make a horizontal implementation by constraining `PT1` with `Tool1Adapter`.

```
PTool1 = Tool1Adapter ‖ PT1
```

`Tool1Adapter` is itself an constraining of `AdapterTool1` with `Tool1` for which we give the definitions below.

```
AdapterTool1 =
     tooladapter-rec(message) .
     tooltb-snd-event(tbterm(message)) .
     tooltb-rec-ack-event(tbterm(message)) .
     tooladapter-snd(ack) .
     AdapterTool1
   + tooladapter-rec(quit) .
     tooltb-snd-event(tbterm(quit))
Tool1 =
     snd(message) .
     rec(ack) .
     Tool1
   + snd(quit)
```

In this constraint, the communication between the actions `tooladapter-rec` and `tooladapter-snd` of `AdapterTool1` and the actions `snd` and `rec` of `Tool1` are enforced.

An implementation for `Component2` can be obtained in a similar way. A generated animation is shown in Figure 4, in which AdapterTool1 just sent a message it had received from Tool1, to ToolBus process PT1.

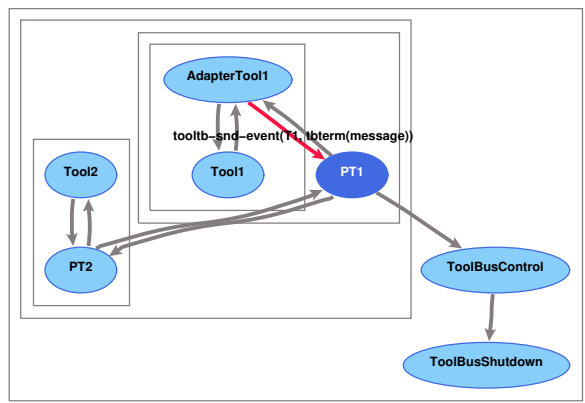

**Figure 4.** Animation of the ToolBus specification example

**Implementation**

The implementation consists of three Tcl/Tk [29] programs (Tool1, its adapter, and Tool2), and a ToolBus script. A screendump of this application at work together with the viewer of the ToolBus is shown in Figure 5. With the viewer it is possible to step through the execution of the ToolBus script and view the variables of the individual processes inside the ToolBus. The ToolBus script is shown below. The `execute` actions in the ToolBus script correspond to starting the adapter for Tool1 and starting Tool2 in parallel with the processes `PT1` and `PT2` respectively.

```
process PT1 is
let
   T1: tool1adapter
in
   execute(tool1adapter, T1?) .
   (
      rec-event(T1, message) .
      snd-msg(t1, t2, message) .
      rec-msg(t2, t1, ack) .
      snd-ack-event(T1, message)
    + rec-event(T1, quit) .
      shutdown("")
   ) * delta
endlet
process PT2 is
let
   T2: tool2
```

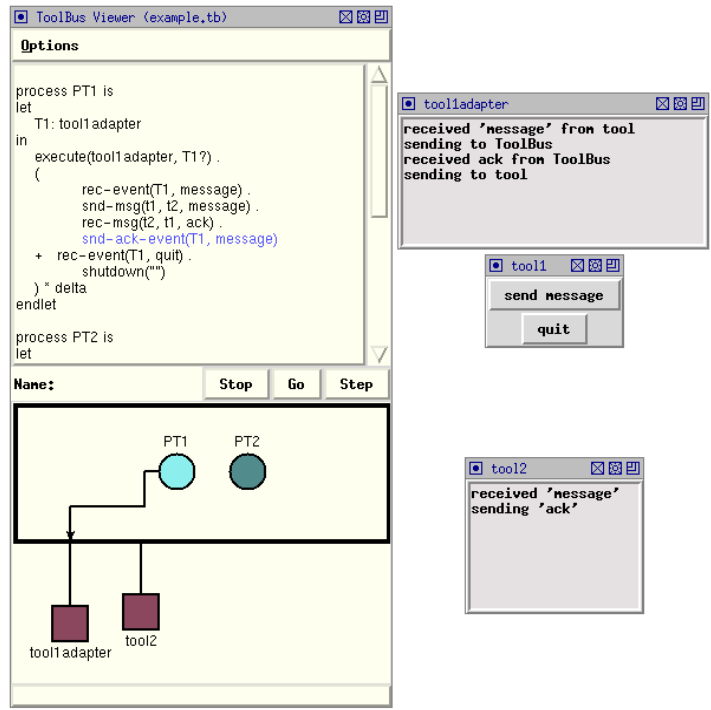

**Figure 5.** Screendump of the example as ToolBus application with viewer

```
in
   execute(tool2, T2?) .
   (
      rec-msg(t1, t2, message) .
      snd-eval(T2, eval(message)) .
      rec-value(T2, value(ack)) .
      snd-msg(t2, t1, ack)
   ) * delta
endlet
tool tool1adapter is { command = "wish-adapter -script tool1adapter.tcl" }
tool tool2 is { command = "wish-adapter -script tool2.tcl" }
toolbus(PT1, PT2)
```

The processes in the ToolBus script use iteration (**\***, where P **\* delta** repeats P infinitely) and the processes in the PSF specification use recursion. In PSF it is also possible to use iteration in this case, since the processes have no arguments to hold the current state. On the other hand, in PSF it is not possible to define variables for storing a global state, so when it is necessary to hold the current state, this must be done through the arguments of a process and be formalized via recursion.

Following the description of the ToolBus processes is the description of how to execute the tools by the execute actions. The last line of the ToolBus script starts the processes `PT1` and `PT2` in parallel.

### 3. Modelling Client/Server Architectures

In this section we investigate the development of implementations based on a client/server architecture from an architecture specification in process algebra. The goal is to develop a software engineering environment for the development of software systems based on a client/server architecture, similar to the PSF-ToolBus software engineering environment described in section 2.

We do this using an application consisting of an operator which can request primitive operations to be performed on some data. We extend this application with basic operations that are build upon primitive operations, and with complex operations build upon basic and primitive operations. The hierarchy of clients and servers is shown in Figure 6.

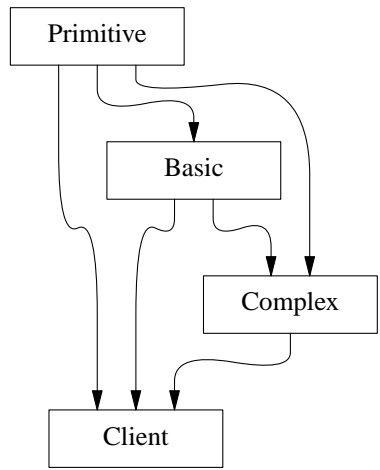

**Figure 6.** Hierachy of clients and servers

*3.1 Architecture Specification*

We start with the specification of the operator in its most simple form. In this form it can input some data, perform a primitive operation, and stop. After stopping, a `snd-quit` is send to the architecture environment in which this specification is put.

```
process module Operator
begin
    exports
    begin
        processes
            Operator
    end
    imports
        ArchitecturePrimitives
    atoms
        input-data
        primitive-operation
        stop
    definitions
        Operator =
            (
                input-data
            + primitive-operation
            + stop .
                snd-quit
            ) * delta
end Operator
```

We refine the `primitive-operation` by adding a sequence of **skip** (the PSF equivalent of $\tau$) actions to it with the use of the algebraic law $a.\tau = a$. These **skip** actions are replaced with `snd` and `rec` actions for communication with another process which provides the services for the primitive operations. Our new architecture is given below with the specification of three modules. The first specifying the necessary data used by the other two.

```
data module ApplicationData
begin
    exports
    begin
        functions
            operator : → ID
            primitive : → ID
            primitive-operation : → DATA
            result : → DATA
    end
    imports
```

```
            ArchitectureTypes
     end ApplicationData
```

For convenience we only give the process definitions representing the operator and the process providing the services.

```
            Operator =
               (
                  input-data
               +  primitive-operation .
                  snd(operator, primitive, primitive-operation) .
                  rec(primitive, operator, result)
               +  stop .
                  snd-quit
               ) * delta

            Primitive =
               (
                  rec(operator, primitive, primitive-operation) .
                  snd(primitive, operator, result)
               ) * delta
```

We extend our architecture specification with basic operations. We add the necessary data to the module ApplicationData and the following alternative sequence of actions to the iteration loop in module Operator.

```
               +  basic-operation .
                  snd(operator, basic, basic-operation) .
                  rec(basic, operator, result)
```

We complete the extension with the addition of module Basic containing the following process definition.

```
            Basic =
               (
                  rec(operator, basic, basic-operation) .
                  snd(basic, operator, result)
               ) * delta
```

A basic operation is build up from primitive operations. Therefor we extend the rec action with the use of the algebraic law $a . \tau = a$ and refine this into the computation of basic operations using services provided by the Primitive process.

```
            Basic =
               (
                  rec(operator, basic, basic-operation) .
                  (
                     (
                        compute-basic .
                        snd(basic, primitive, primitive-operation) .
                        rec(primitive, basic, result)
                     ) *
                     result-basic .
                     snd(basic, c, result)
                  )
               ) * delta
```

The inner loop looks similar to the process Operator with only primitive operations. To serve the Basic process, we extend the Primitive process.

```
            Primitive =
               (
                  rec(operator, primitive, primitive-operation) .
                  snd(primitive, operator, result)
               +  rec(basic, primitive, primitive-operation) .
                  snd(primitive, basic, result)
               ) * delta
```

We can generalize this using a sum construction.

```
            Primitive =
               sum(c in ID,
                  rec(c, primitive, primitive-operation) .
                  snd(primitive, c, result)
               ) * delta
```

In a similar way we can add complex operations that are build up from basic and primitive operations.

```
Complex =
   (
      rec(operator, complex, complex-operation) .
      (
         (
            compute-complex-primitive .
            snd(complex, primitive, primitive-operation) .
            rec(primitive, complex, result)
         + compute-complex-basic .
            snd(complex, basic, basic-operation) .
            rec(basic, complex, result)
         ) *
         result-complex .
         snd(complex, operator, result)
      )
   ) * delta
```

And we generalize process Basic with a sum construction as we did for the process Primitive.

An animation of the complete architecture of the application is shown in Figure 7.

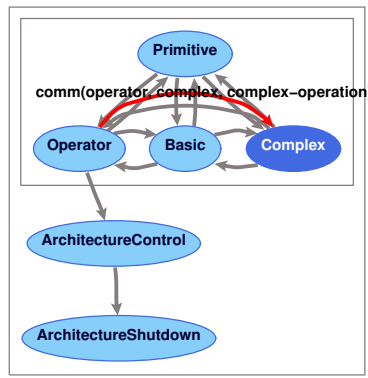

**Figure 7.** Animation of the application architecture

*3.2 Client/Server Architecture Specification*

Going back to our application consisting only of the processes `Operator` and `Primitive` we see that one act as a client and the other as a server. We can hide the fact that the primitive operations are performed by a server through separating the communication with the server from the operator. We specify a client interface.

```
C-I(client, server) =
   (
      sum(s in SERVICE,
         c-rec-call(client, server, s) .
         cs-snd-request(client, server, s)
      ) .
      sum(r in RESULT,
         cs-rec-result(server, client, r) .
         c-snd-return(server, client, r)
      )
   ) * delta
```

The operator can now be specified as a client as follows.

```
C-Operator =
      C-I(operator, primitive)
   ‖  Operator
Operator =
   (
      input-data
   + primitive-operation .
      c-snd-call(primitive, primitive-operation) .
```

```
              c-rec-return(result)
    +   stop .
        snd-quit
    ) * delta
```

The operator as client is the constraining of the client interface `C-I` with the process `Operator`. Communication between these processes takes place through the set of client primitives consisting of the actions `c-snd-call`, `c-rec-call`, `c-snd-return`, and `c-rec-return`.

We can also hide the communication of the server with the client from the execution of the services.

```
        S-I(server) =
            sum(c in ID,
                sum(s in SERVICE,
                    cs-rec-request(c, server, s) .
                    s-snd-call(server, s)
                ) .
                sum(r in RESULT,
                    s-rec-return(server, r) .
                    cs-snd-result(server, c, r)
                )
            ) * delta
```

The primitive server can then be specified as the constraining of the server interface with the process `Primitive` defined as follows.

```
        S-Primitive =
            S-I(primitive)
         || Primitive
        Primitive =
            (
                s-rec-call(primitive-operation) .
                s-snd-return(result)
            ) * delta
```

Communication between these processes takes place through the set of server primitive consisting of the actions `s-snd-call`, `s-rec-call`, `s-snd-return`, and `s-rec-return`.

Our application is formed by combining the processes `C-Operator` and `S-Primitive` which communicate through the set of client/server primitives consisting of the actions `cs-snd-request`, `cs-rec-request`, `cs-snd-result`, and `cs-rec-result`.

We have build a PSF library supporting client/server architecture specifications. This library is similar in setup as the PSF Architecture Library and the PSF ToolBus library. It contains parameterized modules for the client and server interface processes used above. The complete specification of this library can be found in Appendix A.

The processes `Operator` and `Primitive` are the result of applying the following mappings on the processes in the architecture specification.

```
    snd(operator, $1, $2)  →  c-snd-call($1, $2)
    rec($1, operator, result)   →
                               c-rec-return(result)

    rec($1, primitive, primitive-operation)  →
                               s-rec-call(primitive-operation)
    snd(primitive, $1, result)     →
                               s-snd-return(result)
```

Below we give the specification of our application with the use of the PSF Client/Server Architecture Library. The processes `Operator` and `Primitive` are defined in separate modules, which are not shown here.

```
        process module C-Operator
        begin
            exports
            begin
                processes
                    C-Operator
            end
```

```
      imports
         ApplicationData,
         NewC-I {
            Name bound by [
               client → operator,
               server → primitive
            ] to ApplicationData
         },
         Operator
      definitions
         C-Operator =
               C-I(operator, primitive)
            ‖  Operator
   end C-Operator
   process module S-Primitive
   begin
      exports
      begin
         processes
            S-Primitive
      end
      imports
         ApplicationData,
         S-I {
            Name bound by [
               server → primitive
            ] to ApplicationData
         },
         Primitive
      definitions
         S-Primitive =
               S-I(primitive)
            ‖  Primitive
   end S-Primitive
```

We combine the two to form the application system.

```
   process module ApplicationSystem
   begin
      exports
      begin
         processes
            ApplicationSystem
      end
      imports
         NewServer {
            Server bound by [
               Server → S-Primitive
            ] to S-Primitive
            renamed by [
               CS-Server → CS-Primitive
            ]
         },
         NewClient {
            Client bound by [
               Client → C-Operator
            ] to C-Operator
            renamed by [
               CS-Client → CS-Operator
            ]
         }
      definitions
         ApplicationSystem =
               CS-Primitive
            ‖  CS-Operator
   end ApplicationSystem
```

And finally we put the application system in a client/server environment.

```
   process module Application
   begin
      imports
```

```
            ClientServer {
               System bound by [
                  System → ApplicationSystem
               ] to ApplicationSystem
               renamed by [
                  ClientServer → Application
               ]
            }
      end Application
```

We extend the application with a server for the basic operations. First, we add a client interface for communication with the primitive operation server.

```
            process module C-Basic
            begin
               exports
               begin
                  processes
                     C-Basic
               end
               imports
                  ApplicationData,
                  ServerPrimitives,
                  NewC-I {
                     Name bound by [
                        client → basic,
                        server → primitive
                     ] to ApplicationData
                  },
                  Basic
               processes
                  Services
               definitions
                  C-Basic =
                        C-I(basic, primitive)
                     ‖  Basic
            end C-Basic
```

To the resulting client we add a server interface.

```
            process module S-Basic
            begin
               exports
               begin
                  processes
                     S-Basic
               end
               imports
                  S-I {
                     Name bound by [
                        server → basic
                     ] to ApplicationData
                  },
                  NewClient {
                     Client bound by [
                        Client → C-Basic
                     ] to C-Basic
                     renamed by [
                        CS-Client → SC-Basic
                     ]
                  }
               definitions
                  S-Basic =
                        S-I(basic)
                     ‖  SC-Basic
            end S-Basic
```

It is also possible to add the server interface first and then add a client interface to the result. The basic operation server can now be added to `ApplicationSystem` through the import of the library module `NewServer` and binding of its parameter to the process `S-Basic`.

The application can be extended with complex operations in a similar manner as with the basic operations.

However, the server for the complex operations has two client interfaces, one for communication with the basic operations server and one for communication with the primiteve operations server.

An animation of the architecture of the client/server application is shown in Figure 8.

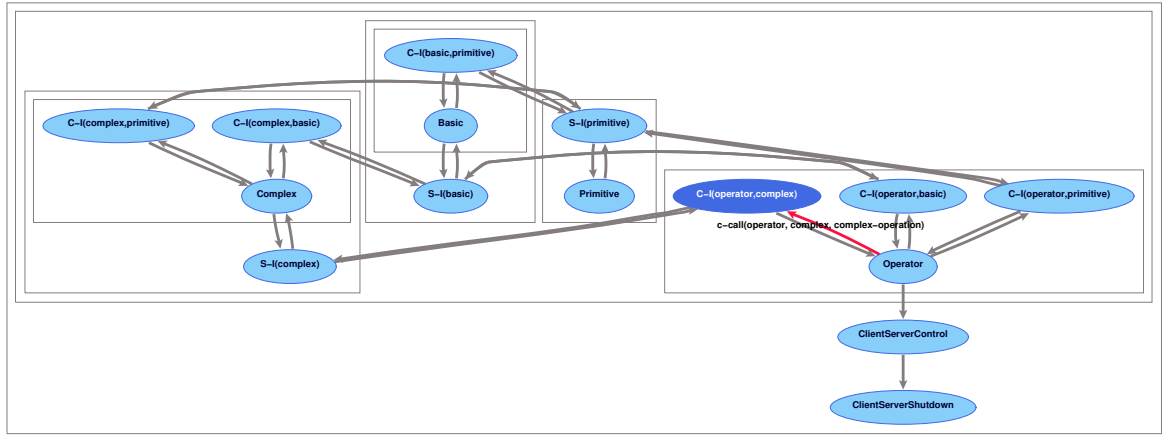

**Figure 8.** Animation of the client/server application architecture

## 4. The PSF-Client/Server Software Engineering Environment

In the previous chapter we developed an architecture specification. From this architecture we developed a specification of client/server application by applying mappings on the processes of the architecture and constraining the resulting processes with client and server interface processes. We can describe the development process as a Software Engineering Environment (SEE) consisting of two workbenches, similar to the PSF-ToolBus SEE described in section 2. This PSF-Client/Server Software Engineering Environment is shown in Figure 9. Objects to be specified are presented as **bold boxes**, workbench tools as ellipses, and generated objects as *slanted boxes*.

The PSF-Client/Server SEE differs from the PSF-ToolBus SEE in that here the tools are constrained with the processes that take care of the communication between the tools instead of the other way round. The reason for this is that there is no choice in how the communications between the clients and the servers take place, making it possible to apply what is called a pattern for the communications. Such a pattern is a generalized structure with parameters that are to be given a value when the pattern is applied.

Another difference is that the processes for constraining are generated from the processes that are constrained. This is possible since we can make use of a pattern for the communications between the clients and the servers. The values for the parameters of the patterns applied are deduced from the type of operations requested by the clients and servers and the knowledge of which operations a particular server can perform.

## 5. Client/Server Architecture based Implementations

From the client/server architecture specification a client/server based implementation can be developed. Such an implementation can be build using almost any implementation language. For a lot of implementation languages an extension package/library already exists that implements the client/server communication, hiding the encoding/decoding of the data used in the communication and hiding the details of the protocol used for the communication.

Through the hiding of the implementation details a request of a client with the receiving of the result of this request looks just like a function call. A server consist of a set functions representing the different services. The calling of these functions is done through a request handler that is hidden from the implementation of the services.

In Appendix B an example of a client/server implementation is given using Perl [32] as implementation language combined with a package that implements web services based on Remote Procedure Calls (RPCs) and the Extensible Markup Language (XML) using the HTTP protocol. For a detailed description of programming web services with Perl we refer to [30].

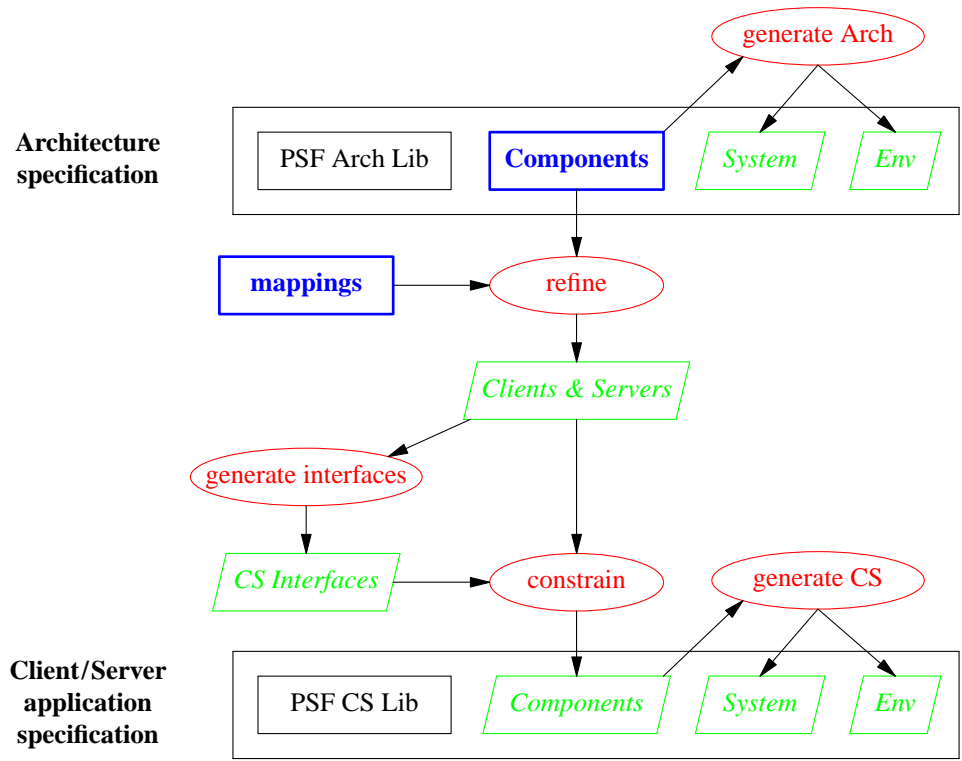

**Figure 9.** The PSF-Client/Server Software Engineering Environment

## 6. Related Work

In the literature several architecture description languages have been proposed and some are based on a process algebra, such as *Wright* [2], *Darwin* [19], and *PADL* [6]. A comparison of several ADL's can be found in [23]. Most of the ADL's do not have any or very little support for refinement. SADL [24][25] however, has been specially designed for supporting architecture refinement. In SADL, different levels of specifications are related by refinement mappings, but the only available tool is a checker. LOTOS [7], a specification language similar to PSF, is used in [18] for the formal description of architectural styles as LOTOS patterns, and in [31] it is used as an ADL for the specification of middleware behaviour.

Formal development techniques such as B [1], VDM [17], and Z [8] provide refinement mechanisms, but they do not have support for architecture descriptions. The $\pi$-Method [26] has been built from scratch to support architecture-centric formal software engineering. It is based on the higher-order typed $\pi$-calculus and mainly built around the architecture description language $\pi$-ADL [27] and the architecture refinement language $\pi$-ARL [28]. Tool support comes in the form of a visual modeler, animator, refiner, and code synthesiser.

Modelling client / server architectures can be done in the above mentioned ADL's and development techniques. Some ADL's provide patterns/styles for clients and server. In contrast to this, we showed that in our approach different types of components and the interaction with these components can be added in a relative easy manner.

To our knowledge there is no work done on generalizing software engineering workbenches and creating software engineering environment from instances of the generalized workbenches. There are many meta software development environments with which an environment can be created by integrating a set of existing tools. Such integration can easily be developed with the PSF-ToolBus software engineering environment as is shown in [11]. Here, an integrated development environment for PSF is created from the tools of the PSF Toolkit using the ToolBus to control the communication between the tools.

**7. Conclusions**

We described how software systems based on a client / server architecture can be developed with process algebra in a similar way as described in previous work for software systems based on the ToolBus. We presented this development process more formally by presenting the tools used in this process in a CASE setting, resulting in the PSF-Client/Server SEE.

The PSF-Client/Server SEE differs from the PSF-ToolBus SEE in that the the processes for constraining can be generated from the processes that describe the clients and servers. This is due to the fact that we use a pattern for the communications between the clients and the servers, and the knowledge of which operations a particular server can perform.

## A. Client/Server Architecture Library

```
data module ClientServerTypes
begin
   exports
   begin
      sorts
         ID,
         SERVICE,
         RESULT
   end
end ClientServerTypes

process module ClientServerPrimitives
begin
   exports
   begin
      atoms
         cs-snd-request : ID # ID # SERVICE
         cs-rec-request : ID # ID # SERVICE
         cs-request : ID # ID # SERVICE
         cs-snd-result : ID # ID # RESULT
         cs-rec-result : ID # ID # RESULT
         cs-result : ID # ID # RESULT
   end
   imports
      ClientServerTypes
   communications
      cs-snd-request(o, d, s) | cs-rec-request(o, d, s) = cs-request(o, d, s)
         for o in ID, d in ID, s in SERVICE
      cs-snd-result(o, d, r) | cs-rec-result(o, d, r) = cs-result(o, d, r)
         for o in ID, d in ID, r in RESULT
end ClientServerPrimitives

process module ServerPrimitives
begin
   exports
   begin
      atoms
         s-snd-call : ID # SERVICE
         s-rec-call : SERVICE
         s-call : ID # SERVICE
         s-snd-return : RESULT
         s-rec-return : ID # RESULT
         s-return : ID # RESULT
   end
   imports
      ClientServerTypes
   communications
      s-snd-call(n, s) | s-rec-call(s) = s-call(n, s)
         for n in ID, s in SERVICE
      s-snd-return(r) | s-rec-return(n, r) = s-return(n, r)
         for n in ID, r in RESULT
end ServerPrimitives

process module S-I
begin
   parameters
      Name
      begin
         functions
            server : → ID
      end Name
   exports
   begin
      processes
         S-I : ID
   end
   imports
      ClientServerPrimitives,
      ServerPrimitives
   variables
      d : → ID
```

```
        definitions
            S-I(server) =
                sum(o in ID,
                    sum(s in SERVICE,
                        cs-rec-request(o, server, s) .
                        s-snd-call(server, s)
                    ) .
                    sum(r in RESULT,
                        s-rec-return(server, r) .
                        cs-snd-result(server, o, r)
                    )
                ) * delta
end S-I

process module NewServer
begin
    parameters
        Server
        begin
            processes
                Server
        end Server
    exports
    begin
        processes
            CS-Server
    end
    imports
        ServerPrimitives
    sets
        of atoms
            ServerH = {
                s-snd-call(n, s), s-rec-call(s),
                s-snd-return(r), s-rec-return(n, r)
                | n in ID, s in SERVICE, r in RESULT
            }
    definitions
        CS-Server =
            encaps(ServerH,
                Server
            )
end NewServer

process module ClientPrimitives
begin
    exports
    begin
        atoms
            c-snd-call : ID # SERVICE
            c-rec-call : ID # ID # SERVICE
            c-call : ID # ID # SERVICE
            c-snd-return : ID # ID # RESULT
            c-rec-return : RESULT
            c-return : ID # ID # RESULT
            snd-quit
    end
    imports
        ClientServerTypes
    communications
        c-snd-call(d, s) | c-rec-call(o, d, s) = c-call(o, d, s)
            for o in ID, d in ID, s in SERVICE
        c-snd-return(o, d, r) | c-rec-return(r) = c-return(o, d, r)
            for o in ID, d in ID, r in RESULT
end ClientPrimitives

process module NewC-I
begin
    parameters
        Name
        begin
            functions
                client : → ID
                server : → ID
```

```
            end Name
      exports
      begin
         processes
            C-I : ID # ID
      end
      imports
         ClientServerPrimitives,
         ClientPrimitives
      variables
         o : → ID
         d : → ID
      definitions
         C-I(client, server) =
            (
               sum(s in SERVICE,
                  c-rec-call(client, server, s) .
                  cs-snd-request(client, server, s)
               ) .
               sum(r in RESULT,
                  cs-rec-result(server, client, r) .
                  c-snd-return(server, client, r)
               )
            ) * delta
end NewC-I

process module NewClient
begin
   parameters
      Client
      begin
         processes
            Client
      end Client
   exports
   begin
      processes
         CS-Client
   end
   imports
      ClientPrimitives
   sets
      of atoms
         ClientH = {
            c-snd-call(d, s), c-rec-call(o, d, s),
            c-snd-return(o, d, r), c-rec-return(r)
            | o in ID, d in ID, s in SERVICE, r in RESULT
         }
   definitions
      CS-Client =
         encaps(ClientH,
            Client
         )
end NewClient

process module ClientServer
begin
   parameters
      System
      begin
         processes
            System
      end System
   exports
   begin
      processes
         ClientServer
   end
   imports
      ClientServerPrimitives
   atoms
      rec-quit
      quit
```

```
            snd-shutdown
            rec-shutdown
            shutdown
     processes
            ClientServerControl
            ClientServerShutdown
     sets
            of atoms
                H = {
                    cs-snd-request(o, d, s), cs-rec-request(o, d, s),
                    cs-snd-result(o, d, r), cs-rec-result(o, d, r)
                    | o in ID, d in ID, s in SERVICE, r in RESULT
                }
                ClientServerH = {
                    snd-quit, rec-quit,
                    snd-shutdown, rec-shutdown
                }
     communications
            snd-quit | rec-quit = quit
            snd-shutdown | rec-shutdown = shutdown
     definitions
            ClientServer =
                    encaps(ClientServerH,
                        disrupt(
                            encaps(H, System),
                            ClientServerShutdown
                        )
                    || ClientServerControl
                    )
            ClientServerControl =
                    rec-quit .
                    snd-shutdown
            ClientServerShutdown = rec-shutdown
end ClientServer
```

## B. An example implementation in Perl

We show how the application specified in section 3 can be implemented using web services. As an example we implement a calculator that can perform the operations on natural numbers. The operations consists of `successor`, `predecessor`, and `iszero` as primitive operations, `add` and `subtract` as basic operators, and `multiply` and `divide` as complex operations. As implementation language we use Perl together with the Frontier-RPC package for implementing web services based on Remote Procedure Calls (RPC) and the Extensible Markup Language (XML) using the HTTP protocol. The package completely hides the XML and HTTP protocol details from the user.

The primitive operation server can be implemented as follows.

```perl
use Frontier::Daemon;
use Frontier::RPC2;
my $d = Frontier::Daemon->new(
   methods => {
      succ => \&succ,
      pred => \&pred,
      iszero => \&iszero,
   },
   LocalAddr => 'localhost',
   LocalPort => 1080,
   ReuseAddr => 1,
);
sub succ {
   my $arg = shift;
   return ++ $arg;
}
sub pred {
   my $arg = shift;
   if (iszero($arg)) {
      return 0;
   } else {
      return -- $arg;
   }
}
sub iszero {
   my $arg = shift;
   return $arg == 0 ? 1 : 0;
}
```

The operator client can be implemented as follows

```perl
use Frontier::Client;
my $primitive = Frontier::Client->new(
    url   => "http://localhost:1080/RPC2",
    debug => 0,
);
my @stack = ();
my @args;
my $r;
sub print_stack {
   my $s;
   $s = join(" ", @stack);
   print "($s)0;
}
while (1) {
   print_stack();
   print "> ";
   # read input
   $_ = <>;
   if ($_ eq "") {
      next;
   }
   # strip input
   if (/\^\s*(.*\S)\s*$/) {
      $_ = $1;
   }
   if (/\^(\d+)$/) { # input natural
      push @stack, $1;
   } elsif ($_ eq "++") { # primitive successor
```

```perl
            if ($#stack < 0) {
               print "not enough arguments0;
            } else {
               $args[0] = pop @stack;
               $r = $primitive->call('succ', @args);
               print "$args[0] ++ = $r0;
               push @stack, $r;
            }
         } elsif ($_ eq "--") {   # primitive predecessor
            if ($#stack < 0) {
               print "not enough arguments0;
            } else {
               $args[0] = pop @stack;
               $r = $primitive->call('pred', @args);
               print "$args[0] -- = $r0;
               push @stack, $r;
            }
         } elsif ($_ eq "=") {    # primitive iszero
            if ($#stack < 0) {
               print "not enough arguments0;
            } else {
               $args[0] = pop @stack;
               $r = $primitive->call('iszero', @args);
               print "$args[0] == 0 = $r0;
               push @stack, $r;
            }
         } elsif ($_ eq "p") {    # pop stack
            pop @stack;
         } elsif ($_ eq "c") {    # clear stack
            @stack = ();
         } elsif ($_ eq "q") {    # stop
            last;
         }
      }
}
```

The basic operations client/server can be implemented as follows.

```perl
       use Frontier::Daemon;
       use Frontier::Client;
       my $primitive = Frontier::Client->new(
           url => "http://localhost:1080/RPC2",
           debug => 0,
       );
       my $d = Frontier::Daemon->new(
          methods => {
              add => \&add,
              subtract => \&subtract,
          },
          LocalAddr => 'localhost',
          LocalPort => 1081,
          ReuseAddr => 1,
       );
       sub add {
           my ($arg1, $arg2) = @_;
           while (! $primitive->call('iszero', $arg2)) {
               $arg1 = $primitive->call('succ', $arg1);
               $arg2 = $primitive->call('pred', $arg2);
           }
           return $arg1;
       }
       sub subtract {
           my ($arg1, $arg2) = @_;
           while (! $primitive->call('iszero', $arg2) &&
                  ! $primitive->call('iszero', $arg1)) {
               $arg1 = $primitive->call('pred', $arg1);
               $arg2 = $primitive->call('pred', $arg2);
           }
           return $arg1;
       }
```

The complex operations client/server can be implemented as follows.

```perl
       use Frontier::Daemon;
```

- 24 -

```perl
use Frontier::Client;
my $primitive = Frontier::Client->new(
    url => "http://localhost:1080/RPC2",
    debug => 0,
);
my $basic = Frontier::Client->new(
    url => "http://localhost:1081/RPC2",
    debug => 0,
);
my $d = Frontier::Daemon->new(
   methods => {
       mul => \&multiply,
       div => \÷,
   },
   LocalAddr => 'localhost',
   LocalPort => 1082,
);
sub multiply {
   my ($arg1, $arg2) = @_;
   my $r;
   $r = 0;
   while (! $primitive->call('iszero', $arg2)) {
       $r = $basic->call('add', ($r, $arg1));
       $arg2 = $primitive->call('pred', $arg2);
   }
   return $r;
}
sub lessthan {
   my $arg1 = shift;
   my $arg2 = shift;
   my $r;
   $r = $basic->call('subtract', $arg2, $arg1);
   $r = $primitive->call('iszero', $r);
   return $r == 0 ? 1 : 0;
}
sub divide {
   my ($arg1, $arg2) = @_;
   my $r = 0;
   if (! $primitive->call('iszero', $arg2)) {
       while (! lessthan($arg1, $arg2)) {
           $arg1 = $basic->call('subtract', ($arg1, $arg2));
           $r = $primitive->call('succ', $r);
       }
   }
   return $r;
}
```